# ADDITIONAL NON-COULOMB ELECTRON-ELECTRON INTERACTION IN MULTIELECTRON SYSTEMS

Kavera V.V.
Krasnodar, Russia

E-mail: vad1809@rambler.ru

**Abstract.** The author suggests an approach based on the separation of total energy of multielectron systems into the semi-classical Coulomb part and the non-classical additional part. This approach allows on the one hand to simplify calculations and on the other hand to see very simple and very interesting relations. These relations were not obvious when using more complex methods. The most interesting aspect is that on all levels of chemical matter - from two-electron atoms up to molecules and crystals - we find additional non-classical electron-electron attraction. The most important practical consequence is the discovery that under certain conditions additional electron-electron attraction can exceed usual electron-electron repulsion. It can result in existence of ordered structures of a new type in a special-way exited matter. PACS: 31.15.bu, 31.15.xt

## CONTENTS

## I. INTRODUCTION

The present work will show that the energy of a multielectron system can be presented as two parts.

One of them takes into account all classical electrostatic interaction inside the given system. In what follows we shall name it semi-classical energy as it submits to Bohr's correspondence principle, i.e. it is saved in the classical calculation. Mathematically, it is obtained from the general analytical solution of the Schrodinger equation. This solution for a multielectron problem becomes possible if we assume that all electron-electron interaction reduces only to mutual shielding of the nucleus charge. This supposition allows us to obtain total separation of variables of electrons. Thus multielectron problem is transformed into a sum of one-electron hydrogen-like problems, each of which has a general analytical solution. Any additional types of interactions are not taken into account apart from those that take place in one-electron atom.

The second part represents the rest that is obtained after a subtraction of the first part from the experiment date. In what follows we shall name it additional energy. It takes into account all others kinds of interactions that were not taken into account in the first part. I.e. it involves all known additional interactions (for example exchange, correlation, etc.), and possible unknown types of interactions that are absent in the case of one-electron atom, but occur in the multielectron case. Analyzing this additional energy is the purpose of the given work.

From the beginning we analyzed additional energy for intra-atomic interactions using the first approximation of the variational method. We shall show on the example of two-electron atoms that this method allows to divide atom energy into semi-classical Coulomb and additional non-Coulomb parts with a very high degree of accuracy.

Then we analyzed inter-atomic interactions on the example of elementary molecules and crystals. Analysis of additional energy for these systems means in fact analysis of known energies of inter-atomic bounds. Obtained relations were discussed from physical and mathematical points of view. Possible practical consequences will also be considered.

## II. INTRA-ATOMIC INTERACTIONS

### A. Semi-classical energy

It is known that the Schrodinger equation and Coulomb's law is enough for the calculation of properties of hydrogen-like one-electron atom in the nonrelativistic case. The solution for atom energy in this case looks like

$$E_H = \frac{Z^2}{n^2} \quad , \tag{1}$$

where $Z$ - charge of the nucleus, and $n$ - principal quantum number. [Hereinafter everywhere in the text, if it is not specially stipulated, Rydberg (Ry) is used as units of energy.]

The analysis shows that the dependence of energy on a principal quantum number

$$E \sim 1/n^2$$

in formula (1) results naturally from the dependence of energy on the distance between interacting particles

$$E \sim 1/r$$

in the classical Coulomb's law because the average value of radius $r$ depends on $n$ as $r \sim n^2$ in a one-electron atom. Moreover, there is a point of view that if the Coulomb's law was not known, it could be obtained from known spectra of hydrogen-like atoms.

If we neglect the electron-electron interaction (in the so-called zero approximation) in the case of helium-like atom, the formula (1) transforms into

$$E_{He0} = \frac{Z^2}{n_1^2} + \frac{Z^2}{n_2^2} \quad , \tag{2}$$

where $n_1$ and $n_2$ - principal quantum numbers of two electrons. I.e. the three-body problem (helium-like atom) is separated into two two-body problem (hydrogen-like atom), each of which has a precise general analytical solution. (Let's note, that hereinafter we use a term "helium-like atom" for a neutral atom of helium, a negative ion of hydrogen and positive helium-like ions of lithium, beryllium etc.)

To improve formula (2) it is necessary to take into account electron-electron interaction. Eventually it is the main problem of calculation of two-electron atoms. The review of approximate methods for the solution of this problem as of 2000 can be found in [1].

The independent-particle model is a widely used approach for the solution of the many-body problem in physics generally and in atomic physics in particular. It is used when describing x-ray and elementary optical spectra of atoms, the periodic system of elements and when formulating the Pauli principle.

The wave function of two-electron atom in this approximation

$$\psi = \psi_1 \cdot \psi_2 \quad , \tag{3}$$

or, if we take into account exchange degeneration

$$\psi = \frac{1}{\sqrt{2}} [\psi_1(1) \cdot \psi_2(2) \pm \psi_1(2) \cdot \psi_2(1)] \quad , \tag{4}$$

where $\psi_1$ and $\psi_2$ - wave functions of separate electrons, (1) and (2) designate all spatial coordinates of first and second electrons, and signs " + "и - " corresponds to singlet and triplet states respectively.

Functions $\psi_1, \psi_2$ and $\psi$ are determined with the help of the variational principle and there are two ways of its use.

The first case corresponds to the Hartree-Fock method. In this case the analytical form $\psi_1$ and $\psi_2$ is not set beforehand. The variational principle results in differential equations. These equations have a difficult solution only in the numerical form. Final wave functions are noted as numerical tables. These tables are difficult to interpret.

The second case corresponds to the variational method. It was applied for the first time to the atom of helium by Kellner. Later it was advanced by Hylleraas (see review [1]). In this case $\psi_1$ and $\psi_2$ are set in a specific analytical form and they contain some varied parameters. The variational principle results in algebraic equations. These equations have a simple solution. Sometimes this solution is possible not only in a numerical, but also in a general view. Final wave functions are noted as simple analytical functions.

The approximation of the variational method is named as the first, second, third etc. depending on the quantity of varied parameters. We use in the given work the one-configuration first approximation of the variational method (further VM$_1$). Only one parameter varies in this approximation - charge of the nucleus. All interaction of electrons is reduced only to mutual shielding of this charge. Thus VM$_1$ is the most simple from all possible methods that use the independent-particle model.

The description of VM$_1$ can be found in many sources (for example in [2] and [16]) and an in-depth account is not required here. Let's note that we take into account a common wave function of atom $\psi$ as (3), i.e. we don't take into account in this approximation exchange effects that don't have classical analogs. Thus we take into account only the electrostatic forces that are described by the classical Coulomb's law and submit to Bohr's correspondence principle, i.e. are saved in the classical calculation.

Wave functions of separate electrons $\psi_1$ and $\psi_2$ differ from hydrogen-like functions only by the replacement in them of the real charge of the nucleus $Z$ into effective charges $Z_{e1}$ and $Z_{e2}$, that are variational parameters of the variational equation

$$\iint \psi H \psi^* dV_1 dV_2 = \min , \tag{5}$$

where $H$ – the non-relativistic two-electron Hamiltonian reads in atomic units ( a.e.)

$$H = \frac{p_1^2 + p_2^2}{2} - \frac{Z}{r_1} - \frac{Z}{r_2} + \frac{1}{r_{12}} , \tag{6}$$

where $r_1$ and $r_2$ are the electron-nucleus distances, $r_{12}$ is the electron-electron distance, whereas $p_1$ and $p_2$ are the individual momenta of the electrons.

Hydrogen-like wave functions can be found in any textbook on quantum mechanics. All terms in (6), except for the last one, are hydrogen-like, and integrals $\iint \psi H \psi^* dV_1 dV_2$ are calculated elementary. The last term in (6) describes electron-electron interaction and, without the consideration of exchange degeneration, its calculation is reduced to the evaluation of the so-called Coulomb integral, that is noted as

$$\iint \psi_1^2 \frac{1}{r_{12}} \psi_2^2 dV_1 dV_2 \quad . \tag{7}$$

The evaluation of similar integrals is explained in [2] and [16].

Eventually the formula of energy (in Ry) of two-electron atom calculated by the method VM$_1$ looks like

$$E_{He1} = \frac{Z_{e1}^2}{n_1^2} + \frac{Z_{e2}^2}{n_2^2} . \tag{8}$$

It is also necessary to take into account the fact that experimental energy of one-electron atoms deviates from formula (1) at the increase of Z because of relativistic effects, influence of the nucleus, etc.

Taking this into account, the formula for the exact value of energy of one-electron atom will transform in

$$E_{H\,exp} = \varepsilon \frac{Z^2}{n^2} \quad , \tag{9}$$

where $\varepsilon$ - a factor that takes into account all deviations from formula (1). In the theory of one-electron atoms $\varepsilon$ can be calculated with absolute precision. In the given work its value is taken from the experiment and is determined from (1) and (9) like

$$\varepsilon = \frac{E_{H\,exp}}{E_H}$$

or

$$\varepsilon = E_{H\,exp} \frac{n^2}{Z^2} \quad , \tag{10}$$

where $E_{H\,exp}$ - experimental value of energy of hydrogen-like atom for given $Z$ and given $n$, and $E_H$ - value calculated by the formula (1).

Formula (8) now will finally look like

$$E_{He1} = \varepsilon_1 \frac{Z_{e1}^2}{n_1^2} + \varepsilon_2 \frac{Z_{e2}^2}{n_2^2} \quad , \tag{11}$$

where $\varepsilon_1$ and $\varepsilon_2$ are determined from (10).

Thus we have presented a three-body problem as a sum of two two-body problems. Each of them has a precise analytical solution and contains only those interactions that take place also in the case of a one-electron atom.

## B. Additional energy

Let's evaluate the accuracy of the approach selected by us. For this purpose we shall define a residual between energy calculated by formula (11) and precision energy taken from experiment.

Though the similar evaluation of accuracy of the method VM$_1$ can be found for a ground state of two-electron atoms in various sources (for example in [2]), but we do not know of any publications of similar research for doubly excited states (DES).

### 1. Intrashell states

Let's first consider $nl_1nl_2$ configurations of helium-like atoms where $n = n_1 = n_2$, and $l_1$ and $l_2$ are orbital quantum numbers. Formula (11) in this case looks as

$$E_{He1} = \frac{\varepsilon}{n^2}(Z_{e1}^2 + Z_{e2}^2) \quad . \tag{12}$$

The values $E_{He1}$, calculated by the VM$_1$ method, values of energy $E_{He\,exp}$, taken from the experiment, and the value of additional energy $E_{add} = E_{He\,exp} - E_{He1}$ are given in a Table I.

**TABLE I. The data of experiment and calculations for $nl_1nl_2$ states of heliumlike atoms.**

| State | Z | $E_{He\,exp}$ (Ry) | $E_{He1}$ (Ry) | $E_{add}$ (Ry) | C | $E_{He}$ (Ry) | Ref. $E_{He\,exp}$ |
|-------|---|-------------------|----------------|----------------|---|---------------|--------------------|
| *1s1s (1S)* | 1 | 1,0550 | 0,9448 | 0,1101 | 0,110 | 1,0559 | [3] |
| | 2 | 5,8068 | 5,6948 | 0,1120 | 0,112 | 5,8059 | [4] |
| | 3 | 14,5597 | 14,4457 | 0,1140 | 0,114 | 14,5568 | [4] |
| | 4 | 27,3131 | 27,1987 | 0,1143 | 0,114 | 27,3099 | [4] |
| | 5 | 44,0699 | 43,9561 | 0,1137 | 0,114 | 44,0672 | [4] |
| | 6 | 64,8318 | 64,7202 | 0,1116 | 0,112 | 64,8313 | [4] |
| | 7 | 89,6035 | 89,4946 | 0,1089 | 0,109 | 89,6057 | [4] |
| | 8 | 118,3845 | 118,2829 | 0,1016 | 0,102 | 118,3940 | [4] |
| | 9 | 151,1885 | 151,0900 | 0,0985 | 0,099 | 151,2011 | [4] |
| | 10 | 188,0111 | 187,9201 | 0,0910 | 0,091 | 188,0312 | [4] |
| *2s2s (1S)* | 1 | 0,2972 | 0,2444 | 0,0528 | 0,106 | 0,2999 | [5] |
| | 2 | 1,5571 | 1,4436 | 0,1135 | 0,114 | 1,5547 | [6] |
| | 3 | 3,8077 | 3,6431 | 0,1646 | 0,110 | 3,8098 | [7] |
| | 4 | 7,0788 | 6,8434 | 0,2354 | 0,118 | 7,0656 | [7] |
| | 5 | 11,3354 | 11,0452 | 0,2902 | 0,116 | 11,3229 | [8] |
| *3s3s (1S)* | 2 | 0,7206 | 0,6431 | 0,0776 | 0,116 | 0,7171 | [9] |
| | 6 | 7,4304 | 7,2260 | 0,2044 | 0,102 | 7,4482 | [10] |
| | 7 | 10,2379 | 9,9849 | 0,2530 | 0,108 | 10,2442 | [11] |
| | 8 | 13,4581 | 13,1900 | 0,2681 | 0,101 | 13,4863 | [10] |
| *2s2p (1P)* | 1 | 0,2520 | 0,2313 | 0,0207 | 0,083 | 0,2498 | [5] |
| | 2 | 1,3860 | 1,4071 | -0,0210 | 0,084 | 1,3700 | [12] |
| | 3 | 3,5121 | 3,5832 | -0,0711 | 0,095 | 3,4906 | [15] |
| | 4 | 6,6378 | 6,7600 | -0,1222 | 0,098 | 6,6119 | [7] |
| | 5 | 10,7842 | 10,9383 | -0,1542 | 0,088 | 10,7346 | [8] |
| *3s3p (1P)* | 1 | 0,1249 | 0,1062 | 0,0187 | 0,112 | 0,1247 | [5] |
| | 2 | 0,6712 | 0,6353 | 0,0359 | 0,108 | 0,6723 | [12] |
| *4s4p (1P)* | 2 | 0,3888 | 0,3595 | 0,0294 | 0,117 | 0,3872 | [12] |
| *5s5p (1P)* | 1 | 0,0491 | 0,0390 | 0,0101 | 0,101 | 0,0501 | [5] |
| | 2 | 0,2528 | 0,2307 | 0,0221 | 0,110 | 0,2529 | [12] |
| *6s6p (1P)* | 1 | 0,0348 | 0,0272 | 0,0076 | 0,091 | 0,0364 | [5] |
| | 2 | 0,1760 | 0,1605 | 0,0156 | 0,094 | 0,1790 | [12] |
| *2s2p (3P)* | 1 | 0,2841 | 0,2313 | 0,0528 | 0,106 | 0,2869 | [5] |
| | 2 | 1,5218 | 1,4071 | 0,1147 | 0,115 | 1,5182 | [6] |
| | 3 | 3,7536 | 3,5832 | 0,1704 | 0,114 | 3,7498 | [7] |
| | 4 | 6,9906 | 6,7600 | 0,2306 | 0,115 | 6,9823 | [7] |
| | 5 | 11,2292 | 10,9383 | 0,2909 | 0,116 | 11,2161 | [13] |
| | 6 | 16,4630 | 16,1188 | 0,3442 | 0,115 | 16,4521 | [13] |
| | 10 | 47,4169 | 46,8875 | 0,5294 | 0,106 | 47,4431 | [13] |
| | 12 | 68,9347 | 68,3223 | 0,6125 | 0,102 | 68,9889 | [13] |
| *3s3p (3P)* | 2 | 0,6988 | 0,6353 | 0,0635 | 0,095 | 0,7094 | [14] |
| | 7 | 10,2085 | 9,9533 | 0,2552 | 0,109 | 10,2126 | [11] |
| *2p2p (1S)* | 2 | 1,2454 | 1,3393 | -0,0939 | 0,094 | 1,1171 | [6] |
| | 3 | 3,2578 | 3,4764 | -0,2186 | 0,109 | 3,1430 | [7] |
| | 4 | 6,2414 | 6,6141 | -0,3727 | 0,124 | 6,1697 | [7] |
| | 5 | 10,2991 | 10,7533 | -0,4542 | 0,114 | 10,1978 | [8] |
| *3p3p (1S)* | 2 | 0,6354 | 0,6246 | 0,0108 | - | 0,6246 | [9] |
| *2p2p (1D)* | 1 | 0,2561 | 0,2026 | 0,0535 | 0,107 | 0,2582 | [5] |
| | 2 | 1,4049 | 1,3393 | 0,0656 | 0,131 | 1,3949 | [6] |
| | 3 | 3,5376 | 3,4764 | 0,0612 | 0,122 | 3,5319 | [7] |
| | 4 | 6,6744 | 6,6141 | 0,0603 | 0,121 | 6,6697 | [7] |
| | 5 | 10,8149 | 10,7533 | 0,0616 | 0,123 | 10,8089 | [13] |
| | 6 | 15,9391 | 15,8947 | 0,0444 | 0,089 | 15,9502 | [13] |
| *3p3p (1D)* | 2 | 0,6949 | 0,6246 | 0,0703 | 0,105 | 0,6987 | [9] |
| | 6 | 7,3503 | 7,1638 | 0,1865 | 0,093 | 7,3860 | [10] |
| | 7 | 10,1497 | 9,9118 | 0,2379 | 0,102 | 10,1711 | [11] |
| | 8 | 13,3589 | 13,1059 | 0,2529 | 0,095 | 13,4022 | [10] |
| *2p2p (3P)* | 2 | 1,4205 | 1,3393 | 0,0812 | 0,122 | 1,4134 | [13] |
| | 4 | 6,7668 | 6,6141 | 0,1527 | 0,115 | 6,7623 | [13] |
| | 5 | 10,9450 | 10,7533 | 0,1917 | 0,115 | 10,9385 | [13] |
| | 6 | 16,1096 | 15,8947 | 0,2149 | 0,107 | 16,1169 | [13] |
| | 8 | 29,5400 | 29,1887 | 0,3514 | 0,132 | 29,4849 | [13] |
| *3p3p (3P)* | 7 | 10,0615 | 9,9118 | 0,1497 | 0,128 | 10,0415 | [11] |

| | | | | | | |
|---|---|---|---|---|---|---|
| **3d3d (1D)** | 2 | 0,6545 | 0,5780 | 0,0765 | 0,115 | 0,6520 | [9] |
| | 6 | 7,2040 | 7,0037 | 0,2004 | 0,100 | 7,2259 | [10] |
| | 7 | 10,0101 | 9,7233 | 0,2868 | 0,123 | 9,9826 | [11] |
| | 8 | 13,2090 | 12,8890 | 0,3200 | 0,120 | 13,1853 | [10] |
| **3d3d (1G)** | 2 | 0,6133 | 0,5779 | 0,0353 | 0,106 | 0,6150 | [14] |

The analysis of dependence of additional energy on $Z$ and $n$ shows that the $nl_1nl_2$ states are divided into two groups. The first group involves the lowest states for the given configuration, i.e. states with the lowest possible $n$. They are states in which at least one electron is on the orbit for which $n = l + 1$. In the old quantum theory this condition corresponds to the special case of circular orbits. The second group involves all remaining states of configurations, i.e. states with $n > l + 1$ for everyone from two electrons. The formulas were obtained

$$E_{add} = \frac{C}{2} \cdot \frac{Z}{n}[k_1 - k_2 \frac{(Z-1)}{Z}]$$ (13)

and

$$E_{add} = \frac{C}{2} \cdot \frac{Z}{n}k_1$$ (14)

respectively for the first and the second group of states, where $k_1$ and $k_2$ are integer factors, the values of which are given in Table II, and $C$ is the numerical factor. Values $C$, at which formulas (13) and (14) give a precise coincidence with experiment, are given in Table I. It is easy to see, taking into account errors of measurements, that $C$ is a constant, the average value of which is close to 1/9.

It is possible to combine formulas (13) and (14) for example like

$$E_{add} = \frac{C}{2} \cdot \frac{Z}{n}[k_1 - k_2 \frac{(Z-1)}{Z} \cdot e^{-m \cdot n_{r1} \cdot n_{r2}}] \quad ,$$

where $n_{r1} = n - l_1 - 1$ and $n_{r2} = n - l_2 - 1$ are the so-called radial quantum numbers of electrons, and the unknown factor $m$ should have a very large, but not infinite, value for the exponential factor to transform into 1, when $n_r = 0$ for at least one electron, and decreased to practically 0 already at $n_r = 1$ (and more) for each of electrons.

**TABLE II. Factors $k_1$ and $k_2$.**

| State | $k_1$ | $k_2$ |
|---|---|---|
| **nsns(1S)** | 2 | 2 |
| **npnp(1D)** | 2 | 2 |
| **nsnp(3P)** | 2 | 0 |
| **npnp(3P)** | 1 | 0 |
| **nsnp(1P)** | 1 | 3 |
| **npnp(1S)** | 0 | 4 |

In the specific case of states *nsns (1S)*, *npnp (1D)*, *ndnd (1G)* etc., when both electrons are on the same orbit, or, in other words, occupy the same quantum cell, formulas (13) and (14) become especially simple

$$E_{add} = C\frac{1}{n}$$ (15)

for $n = l + 1$ and

$$E_{add} = C\frac{Z}{n}$$ (16)

for $n > l + 1$.

Let's note that for these states the result of calculation by a method $VM_1$ does not depend on the fact whether we use a common wave function of atom as (3) or (4), i.e. whether we take into account exchange degeneration or not.

The particular case of formula (15) for $n = 1$ corresponds to a ground state of helium-like atoms (1s1s) and leads to $E_{add} = C$, i.e. the additional energy in this case does not depend on $Z$ and it was noted by Bethe as a curious fact in [2]. This fact (and this is only a particular case of all states considered by us) attracted attention of other researchers in 1930s – 1960s. The main difference of the present work from those works is the application of the same well-known but a little forgotten method $VM_1$ to the newest experimental data of doubly excited states. If these data were known in 1930s – 1960s the $VM_1$ method wouldn't have been forgotten, and formulas (13) - (16), would've been obtained back then. However it was practically impossible to obtain these formulas without complete experimental data at least for $n = 3$ for $nl_1nl_2$ states of helium. For the first time these data appeared in [14] in 1997 and then in [9] in 2001. Thus only from this time it was possible to try to determine the dependence of additional energy $E_{add}$ on $Z$ and $n$ in case of the $VM_1$ method.

Let's note that $VM_1$ is not one of many methods, but it is unique. Only this method allows us to take into account the electron-electron interaction and at the same time completely to divide their variables. Thus the final result is obtained in a simple analytical form. It is very important that $VM_1$ bases on a maximum simple and absolute clear physical model. Wave functions of electrons do not contain any additional terms in comparison to hydrogen-like functions. As opposed to this, all other more complex approximations of the variational method (and other methods) firstly do not allow completely to divide variables, and secondly introduce more and more additional terms into the wave function. The physical sense of these additional terms is not explained at all and this problem interests nobody. To the present time the progress of computers has allowed us to use several hundreds and even thousands additional variational parameters in some calculations. The only justification of these additional terms is better convergence of the calculation with the experiment. However in reality it means adjustment of the calculation to the initially known experimental data. It greatly reminds of the system of Ptolemy and its epicycles. A relatively big number of epicycles allows to describe motion of planets very precisely even without knowledge of the Kepler's Laws and Newton's Laws. Certainly the complexity of calculations is much higher in this case and it was the main stimulus for discovery of a more simple heliocentric system. However it is possible to assume if computers had already existed in the times of Copernicus, this stimulus could not have appeared and we would have used epicycles until now and would have still thought that the Sun and the planets rotate round the Earth.

Actually though $VM_1$ does not give a total convergence with the experiment, but its residual with the experiment is described by so simple formulas (see above) that there is a question, whether it is an accident or there are some physical properties in atoms (and generally - in multielectron systems) that had been escaping the researchers' attention?

In any case it is already possible now to offer a very simple and precise semi-empirical algorithm of calculation that allows to describe known lines of spectra of helium-like atoms and to predict or to help to identify unknown lines.

This algorithm for calculation of total energy of $nl_1nl_2$ states looks like

$$E_{He} = E_{He1} + E_{add} \ ,$$ (17)

where $E_{He1}$ is calculated by the $VM_1$ method and results from formula (12), and $E_{add}$ results from formulas (13) and (14) and is entered from the analysis of the experimental data.

In contrast to the more complex approximations of a variational method (and other methods) mentioned above, the convergence with the experiment is reached here by adding only one additional term that has a very simple form. Besides in contrast to a more conventional point of view, we assume that this term originates not from the mathematical complexity of the problem, but from the existence of additional physical interactions.

Let's also note that from the mid 1960s (see [1]) the point of view that the independent-particle model is inapplicable for describing doubly excited states has prevailed. This conclusion is based on the fact that the Hartree-Fock method was helpless for this class of states. The obtained above formulas show that at least in case of the VM$_1$ method the independent-particle model (i.e. single-particle method) can be applied successfully to doubly excited states as well as to ground states. There is no gap between ground states and doubly excited states in case of VM$_1$.

The data of calculation of energy $E_{He}$ under formula (17) under the supposition that $C = 1/9$, are given in Table I for those of $nl_1nl_2$ states, for which the experimental data are known. Taking into consideration errors of measurements and approximations of calculations, the accordance of calculations and experiment are quite satisfactory. The data of the same calculation for an unknown for today $nsns$ (1S) states at $n$ from 4 up to 10 and $Z$ from 1 up to 10 are given as an example in Table III.

**TABLE III. The data of calculation for *nsns (1S)* states for *n* from 4 up to 10.**

| Z | $E_{He}$(Ry) | | | | | | |
|---|---|---|---|---|---|---|---|
| | *4s4s* | *5s5s* | *6s6s* | *7s7s* | *8s8s* | *9s9s* | *10s10s* |
| 1 | 0,0893 | 0,0616 | 0,0459 | 0,0360 | 0,0293 | 0,0245 | 0,0210 |
| 2 | 0,4175 | 0,2762 | 0,1980 | 0,1500 | 0,1183 | 0,0962 | 0,0802 |
| 3 | 0,9958 | 0,6508 | 0,4612 | 0,3457 | 0,2699 | 0,2174 | 0,1794 |
| 4 | 1,8242 | 1,1854 | 0,8356 | 0,6230 | 0,4839 | 0,3879 | 0,3186 |
| 5 | 2,9029 | 1,8802 | 1,3212 | 0,9820 | 0,7605 | 0,6078 | 0,4979 |
| 6 | 4,2319 | 2,7352 | 1,9180 | 1,4227 | 1,0997 | 0,8771 | 0,7170 |
| 7 | 5,8114 | 3,7504 | 2,6260 | 1,9452 | 1,5014 | 1,1959 | 0,9763 |
| 8 | 7,6417 | 4,9261 | 3,4455 | 2,5494 | 1,9657 | 1,5641 | 1,2758 |
| 9 | 9,7230 | 6,2623 | 4,3763 | 3,2355 | 2,4927 | 1,9818 | 1,6152 |
| 10 | 12,0556 | 7,7592 | 5,4183 | 4,0035 | 3,0823 | 2,4490 | 1,9943 |

## *2. Intershell states*

We shall consider here only $1snl$ state because there are most complete data for them from $n_1l_1n_2l_2$ configurations. Formula (11) in this case looks like

$$E_{Hel} = \varepsilon_1 \cdot Z_{e1}^2 + \varepsilon_2 \frac{Z_{e2}^2}{n^2} \quad , \qquad (18)$$

where the first term corresponds to an internal s-electron in a ground state, and the second term corresponds to an external exited electron.

By analogy to Table I the values $E_{Hel}$, calculated by the VM$_1$ method, values of energy $E_{He\,exp}$, taken from [4], and the value of additional energy $E_{add} = E_{He\,exp} - E_{Hel}$ are given in a Table IV.

**TABLE IV. The data of experiment and calculations for *1snl* states of heliumlike atoms.**

| Z | $E_{He\,exp}$ (Ry) | $E_{Hel}$ (Ry) | $E_{add}$ (Ry) | $E_{He}$ (Ry) | Z | $E_{He\,exp}$ (Ry) | $E_{Hel}$ (Ry) | $E_{add}$ (Ry) | $E_{He}$ (Ry) |
|---|---|---|---|---|---|---|---|---|---|
| | *1s2s (1S)* | | | | | *1s2s (3S)* | | | |
| 2 | 4,2915 | 4,2919 | -0,0004 | 4,2919 | 2 | 4,3501 | 4,2921 | 0,0580 | 4,3476 |
| 3 | 10,0819 | 10,1225 | -0,0405 | 10,0808 | 3 | 10,2217 | 10,1230 | 0,0987 | 10,2202 |
| 4 | 18,3719 | 18,4543 | -0,0825 | 18,3710 | 4 | 18,5968 | 18,4555 | 0,1413 | 18,5944 |
| 5 | 29,1641 | 29,2890 | -0,1249 | 29,1640 | 5 | 29,4755 | 29,2909 | 0,1846 | 29,4714 |
| 6 | 42,4600 | 42,6279 | -0,1679 | 42,4612 | 6 | 42,8587 | 42,6308 | 0,2279 | 42,8531 |
| 7 | 58,2625 | 58,4737 | -0,2111 | 58,2653 | 7 | 58,7490 | 58,4778 | 0,2712 | 58,7417 |
| 8 | 76,5721 | 76,8286 | -0,2566 | 76,5786 | 8 | 77,1529 | 76,8342 | 0,3187 | 77,1397 |

| 9 | 97,3957 | 97,6961 | -0,3005 | 97,4045 | 9 | 98,0615 | 97,7033 | 0,3581 | 98,0505 |
| 10 | 120,7352 | 121,0787 | -0,3434 | 120,7453 | 10 | 121,4892 | 121,0877 | 0,4015 | 121,4766 |
| | *1s3s (1S)* | | | | | *1s3s (3S)* | | | |
| 2 | 4,1222 | 4,1212 | 0,0010 | 4,1212 | 2 | 4,1370 | 4,1212 | 0,0158 | 4,1376 |
| 3 | 9,4677 | 9,4784 | -0,0107 | 9,4660 | 3 | 9,5044 | 9,4784 | 0,0260 | 9,5072 |
| 4 | 17,0366 | 17,0591 | -0,0225 | 17,0344 | 4 | 17,0961 | 17,0591 | 0,0370 | 17,1003 |
| 5 | 26,8306 | 26,8646 | -0,0340 | 26,8276 | 5 | 26,9131 | 26,8646 | 0,0485 | 26,9181 |
| 6 | 38,8500 | 38,8962 | -0,0463 | 38,8469 | 6 | 38,9557 | 38,8962 | 0,0595 | 38,9621 |
| 7 | 53,0981 | 53,1561 | -0,0580 | 53,0944 | 7 | 53,2270 | 53,1561 | 0,0709 | 53,2343 |
| 8 | 69,5732 | 69,6462 | -0,0731 | 69,5722 | 8 | 69,7336 | 69,6462 | 0,0873 | 69,7368 |
| 9 | 88,2813 | 88,3693 | -0,0881 | 88,2829 | 9 | 88,4469 | 88,3693 | 0,0776 | 88,4722 |
| 10 | 109,2330 | 109,3279 | -0,0948 | 109,2291 | 10 | 109,4325 | 109,3279 | 0,1046 | 109,4431 |
| | *1s4s (1S)* | | | | | *1s4s (3S)* | | | |
| 2 | 4,0668 | 4,0661 | 0,0007 | 4,0661 | 2 | 4,0726 | 4,0661 | 0,0065 | 4,0731 |
| 3 | 9,2598 | 9,2640 | -0,0042 | 9,2588 | 3 | 9,2745 | 9,2640 | 0,0105 | 9,2761 |
| 4 | 16,5783 | 16,5881 | -0,0098 | 16,5776 | 4 | 16,6030 | 16,5881 | 0,0149 | 16,6054 |
| 5 | | 26,0396 | | 26,0239 | 5 | | 26,0396 | | 26,0621 |
| 6 | 37,6009 | 37,6198 | -0,0189 | 37,5990 | 6 | 37,6438 | 37,6198 | 0,0240 | 37,6476 |
| 7 | 51,3073 | 51,3307 | -0,0234 | 51,3047 | 7 | 51,3594 | 51,3307 | 0,0287 | 51,3637 |
| 8 | 67,1430 | 67,1743 | -0,0313 | 67,1431 | 8 | 67,2046 | 67,1743 | 0,0303 | 67,2125 |
| 9 | | 85,1531 | | 85,1166 | 9 | | 85,1531 | | 85,1965 |
| 10 | 105,2310 | 105,2694 | -0,0384 | 105,2277 | 10 | 105,3118 | 105,2694 | 0,0424 | 105,3180 |
| | *1s5s (1S)* | | | | | *1s5s (3S)* | | | |
| 2 | 4,0420 | 4,0416 | 0,0004 | 4,0416 | 2 | 4,0449 | 4,0416 | 0,0033 | 4,0451 |
| 3 | 9,1651 | 9,1671 | -0,0021 | 9,1645 | 3 | 9,1724 | 9,1671 | 0,0053 | 9,1734 |
| 4 | | 16,3739 | | 16,3686 | 4 | 16,3810 | 16,3739 | 0,0071 | 16,3828 |
| 5 | | 25,6631 | | 25,6551 | 5 | | 25,6631 | | 25,6746 |
| 6 | 37,0265 | 37,0359 | -0,0094 | 37,0252 | 6 | 37,0479 | 37,0359 | 0,0120 | 37,0501 |
| 7 | 50,4826 | 50,4943 | -0,0117 | 50,4810 | 7 | 50,5088 | 50,4943 | 0,0145 | 50,5112 |
| 8 | | 66,0402 | | 66,0242 | 8 | 66,0540 | 66,0402 | 0,0137 | 66,0598 |
| 9 | | 83,6761 | | 83,6574 | 9 | | 83,6761 | | 83,6983 |
| 10 | 103,3847 | 103,4042 | -0,0195 | 103,3829 | 10 | 103,4254 | 103,4042 | 0,0212 | 103,4291 |
| | *1s6s (1S)* | | | | | *1s6s (3S)* | | | |
| 2 | 4,0288 | 4,0285 | 0,0003 | 4,0285 | 2 | 4,0304 | 4,0285 | 0,0019 | 4,0305 |
| 3 | 9,1141 | 9,1153 | -0,0012 | 9,1137 | 3 | 9,1183 | 9,1153 | 0,0030 | 9,1189 |
| 4 | | 16,2587 | | 16,2556 | 4 | | 16,2588 | | 16,2640 |
| 5 | | 25,4603 | | 25,4557 | 5 | | 25,4603 | | 25,4670 |
| 6 | | 36,7210 | | 36,7148 | 6 | 36,7278 | 36,7210 | 0,0068 | 36,7292 |
| 7 | | 50,0420 | | 50,0342 | 7 | | 50,0427 | | 50,0525 |
| 8 | | 65,4262 | | 65,4170 | 8 | | 65,4274 | | 65,4388 |
| 9 | | 82,8758 | | 82,8650 | 9 | | 82,8776 | | 82,8905 |
| 10 | | 102,3953 | | 102,3829 | 10 | | 102,3953 | | 102,4097 |
| | *1s2p (1P)* | | | | | *1s2p (3P)* | | | |
| 2 | 4,2473 | 4,2312 | 0,0161 | 4,2451 | 2 | 4,2659 | 4,2312 | 0,0348 | 4,2728 |
| 3 | 9,9869 | 9,9963 | -0,0094 | 9,9824 | 3 | 10,0556 | 9,9963 | 0,0593 | 10,0657 |
| 4 | 18,2235 | 18,2628 | -0,0393 | 18,2212 | 4 | 18,3519 | 18,2628 | 0,0891 | 18,3601 |
| 5 | 28,9613 | 29,0323 | -0,0711 | 28,9629 | 5 | 29,1530 | 29,0323 | 0,1207 | 29,1573 |
| 6 | 42,2017 | 42,3064 | -0,1047 | 42,2091 | 6 | 42,4587 | 42,3064 | 0,1523 | 42,4591 |
| 7 | 57,9480 | 58,0873 | -0,1393 | 57,9623 | 7 | 58,2714 | 58,0873 | 0,1840 | 58,2679 |
| 8 | 76,2001 | 76,3777 | -0,1776 | 76,2249 | 8 | 76,5972 | 76,3777 | 0,2195 | 76,5860 |
| 9 | 96,9692 | 97,1808 | -0,2116 | 97,0002 | 9 | 97,4185 | 97,1808 | 0,2377 | 97,4169 |
| 10 | 120,2442 | 120,4990 | -0,2548 | 120,2907 | 10 | 120,7761 | 120,4990 | 0,2770 | 120,7629 |
| | *1s3p (1P)* | | | | | *1s3p (3P)* | | | |
| 2 | 4,1099 | 4,1052 | 0,0047 | 4,1093 | 2 | 4,1158 | 4,1052 | 0,0106 | 4,1175 |
| 3 | 9,4406 | 9,4437 | -0,0031 | 9,4396 | 3 | 9,4611 | 9,4437 | 0,0174 | 9,4643 |
| 4 | 16,9940 | 17,0058 | -0,0118 | 16,9934 | 4 | 17,0312 | 17,0058 | 0,0254 | 17,0346 |
| 5 | 26,7734 | 26,7926 | -0,0191 | 26,7720 | 5 | 26,8268 | 26,7926 | 0,0342 | 26,8296 |
| 6 | 38,7753 | 38,8055 | -0,0302 | 38,7767 | 6 | 38,8479 | 38,8055 | 0,0424 | 38,8508 |
| 7 | 53,0035 | 53,0467 | -0,0432 | 53,0096 | 7 | 53,0967 | 53,0467 | 0,0500 | 53,1002 |
| 8 | 69,4626 | 69,5181 | -0,0555 | 69,4728 | 8 | 69,5800 | 69,5181 | 0,0619 | 69,5798 |
| 9 | 88,1599 | 88,2225 | -0,0626 | 88,1690 | 9 | 88,2877 | 88,2225 | 0,0653 | 88,2924 |
| 10 | 109,0904 | 109,1623 | -0,0718 | 109,1005 | 10 | 109,2374 | 109,1623 | 0,0752 | 109,2404 |
| **1s4p (1P)** | | | | | **1s4p (3P)** | | | | |
| 2 | 4,0618 | 4,0598 | 0,0020 | 4,0615 | 2 | 4,0643 | 4,0598 | 0,0045 | 4,0650 |
| 3 | 9,2485 | 9,2498 | -0,0013 | 9,2481 | 3 | 9,2571 | 9,2498 | 0,0073 | 9,2585 |
| 4 | 16,5616 | 16,5662 | -0,0045 | 16,5609 | 4 | 16,5767 | 16,5662 | 0,0106 | 16,5783 |
| 5 | 26,0016 | 26,0099 | -0,0083 | 26,0012 | 5 | 26,0242 | 26,0099 | 0,0143 | 26,0255 |
| 6 | 37,5698 | 37,5824 | -0,0126 | 37,5702 | 6 | 37,5999 | 37,5824 | 0,0175 | 37,6015 |
| 7 | 51,2682 | 51,2855 | -0,0173 | 51,2699 | 7 | 51,3064 | 51,2855 | 0,0209 | 51,3081 |
| 8 | 67,0974 | 67,1213 | -0,0239 | 67,1022 | 8 | 67,1476 | 67,1213 | 0,0262 | 67,1474 |
| 9 | 85,0636 | 85,0923 | -0,0288 | 85,0698 | 9 | 85,1181 | 85,0923 | 0,0257 | 85,1218 |
| 10 | 105,1715 | 105,2008 | -0,0294 | 105,1748 | 10 | 105,2317 | 105,2008 | 0,0309 | 105,2338 |
| **1s5p (1P)** | | | | | **1s5p (3P)** | | | | |
| 2 | 4,0394 | 4,0384 | 0,0010 | 4,0393 | 2 | 4,0407 | 4,0384 | 0,0023 | 4,0411 |
| 3 | 9,1592 | 9,1600 | -0,0008 | 9,1591 | 3 | 9,1638 | 9,1600 | 0,0037 | 9,1645 |
| 4 | 16,3606 | 16,3629 | -0,0023 | 16,3602 | 4 | 16,3682 | 16,3629 | 0,0053 | 16,3691 |
| 5 | 25,6436 | 25,6481 | -0,0044 | 25,6436 | 5 | 25,6556 | 25,6481 | 0,0075 | 25,6561 |
| 6 | 37,0104 | 37,0169 | -0,0065 | 37,0107 | 6 | 37,0260 | 37,0169 | 0,0091 | 37,0267 |
| 7 | 50,4546 | 50,4714 | -0,0168 | 50,4634 | 7 | 50,4821 | 50,4714 | 0,0107 | 50,4830 |
| 8 | 66,0007 | 66,0134 | -0,0126 | 66,0036 | 8 | 66,0282 | 66,0134 | 0,0148 | 66,0267 |
| 9 | 83,6271 | 83,6453 | -0,0181 | 83,6337 | 9 | 83,6611 | 83,6453 | 0,0159 | 83,6604 |
| 10 | 103,3546 | 103,3694 | -0,0149 | 103,3561 | 10 | 103,3847 | 103,3694 | 0,0153 | 103,3863 |
| **1s6p (1P)** | | | | | **1s6p (3P)** | | | | |
| 2 | 4,0273 | 4,0267 | 0,0006 | 4,0272 | 2 | 4,0280 | 4,0267 | 0,0013 | 4,0282 |
| 3 | 9,1118 | 9,1112 | 0,0006 | 9,1107 | 3 | 9,1134 | 9,1112 | 0,0022 | 9,1138 |
| 4 | 16,2509 | 16,2524 | -0,0015 | 16,2508 | 4 | | 16,2525 | | 16,2561 |
| 5 | 25,4494 | 25,4517 | -0,0023 | 25,4491 | 5 | | 25,4517 | | 25,4563 |
| 6 | 36,7061 | 36,7101 | -0,0040 | 36,7065 | 6 | 36,7153 | 36,7101 | 0,0052 | 36,7157 |
| 7 | 50,0245 | 50,0288 | -0,0043 | 50,0242 | 7 | 50,0330 | 50,0296 | 0,0034 | 50,0362 |
| 8 | 65,4039 | 65,4108 | -0,0069 | 65,4051 | 8 | 65,4241 | 65,4120 | 0,0121 | 65,4197 |
| 9 | 82,8468 | 82,8581 | -0,0112 | 82,8514 | 9 | 82,8681 | 82,8599 | 0,0082 | 82,8686 |
| 10 | 102,3866 | 102,3753 | 0,0113 | 102,3675 | 10 | 102,3976 | 102,3753 | 0,0224 | 102,3850 |
| **1s3d (1D)** | | | | | **1s3d(3D)** | | | | |
| 2 | 4,1109 | 4,1101 | 0,0007 | 4,1101 | 2 | 4,1109 | 4,1101 | 0,0007 | 4,1101 |
| 3 | 9,4450 | 9,4443 | 0,0007 | 9,4443 | 3 | 9,4453 | 9,4443 | 0,0010 | 9,4443 |
| 4 | 17,0024 | 17,0019 | 0,0005 | 17,0019 | 4 | 17,0031 | 17,0019 | 0,0012 | 17,0019 |
| 5 | 26,7846 | 26,7844 | 0,0003 | 26,7844 | 5 | 26,7861 | 26,7844 | 0,0017 | 26,7844 |
| 6 | 38,7921 | 38,7929 | -0,0008 | 38,7929 | 6 | 38,7941 | 38,7929 | 0,0012 | 38,7929 |
| 7 | 53,0273 | 53,0296 | -0,0023 | 53,0296 | 7 | 53,0302 | 53,0296 | 0,0005 | 53,0296 |
| 8 | 69,4907 | 69,4967 | -0,0060 | 69,4967 | 8 | 69,5007 | 69,4967 | 0,0041 | 69,4967 |
| 9 | 88,1662 | 88,1966 | -0,0305 | 88,1966 | 9 | 88,1932 | 88,1966 | -0,0034 | 88,1966 |
| 10 | 109,0734 | 109,1320 | -0,0586 | 109,1320 | 10 | 109,1227 | 109,1320 | -0,0093 | 109,1320 |
| **1s4d (1D)** | | | | | **1s4d (3D)** | | | | |
| 2 | 4,0622 | 4,0613 | 0,0009 | 4,0613 | 2 | 4,0622 | 4,0613 | 0,0009 | 4,0613 |
| 3 | 9,2504 | 9,2490 | 0,0014 | 9,2490 | 3 | 9,2505 | 9,2490 | 0,0015 | 9,2490 |
| 4 | 16,5647 | 16,5629 | 0,0018 | 16,5629 | 4 | 16,5647 | 16,5629 | 0,0018 | 16,5629 |
| 5 | 26,0065 | 26,0043 | 0,0023 | 26,0043 | 5 | 26,0074 | 26,0043 | 0,0031 | 26,0043 |
| 6 | 37,5765 | 37,5744 | 0,0021 | 37,5744 | 6 | 37,5779 | 37,5744 | 0,0035 | 37,5744 |
| 7 | 51,2772 | 51,2752 | 0,0020 | 51,2752 | 7 | 51,2772 | 51,2752 | 0,0021 | 51,2752 |
| 8 | 67,1105 | 67,1086 | 0,0019 | 67,1086 | 8 | 67,1169 | 67,1086 | 0,0084 | 67,1086 |
| 9 | 85,0830 | 85,0772 | 0,0058 | 85,0772 | 9 | 85,0673 | 85,0772 | -0,0099 | 85,0772 |
| 10 | | 105,1833 | | 105,1833 | 10 | | 105,1833 | | 105,1833 |
| **1s4f (1F)** | | | | | **1s4f (3F)** | | | | |
| 2 | 4,0621 | 4,0621 | 0,0000 | 4,0621 | 2 | 4,0621 | 4,0621 | 0,0000 | 4,0621 |

| 3 | 9,2502 | 9,2502 | 0,0000 | 9,2502 | 3 | 9,2502 | 9,2502 | 0,0000 | 9,2502 |
|---|---|---|---|---|---|---|---|---|---|
| 4 | 16,5645 | 16,5646 | 0,0000 | 16,5646 | 4 | 16,5645 | 16,5646 | -0,0001 | 16,5646 |
| 5 | 26,0064 | 26,0064 | 0,0000 | 26,0064 | 5 | 26,0062 | 26,0064 | -0,0001 | 26,0064 |
| 6 | 37,5765 | 37,5768 | -0,0003 | 37,5768 | 6 | 37,5765 | 37,5768 | -0,0003 | 37,5768 |
| 7 |  | 51,2780 |  | 51,2780 | 7 |  | 51,2780 |  | 51,2780 |
| 8 | 67,1089 | 67,1119 | -0,0030 | 67,1119 | 8 | 67,1147 | 67,1119 | 0,0028 | 67,1119 |
| 9 | 85,0555 | 85,0809 | -0,0255 | 85,0809 | 9 | 85,0601 | 85,0809 | -0,0208 | 85,0809 |
| 10 | 105,1368 | 105,1875 | -0,0507 | 105,1875 | 10 | 105,1787 | 105,1875 | -0,0087 | 105,1875 |

The analysis of the data of Table IV results in the formula for additional energy

$$E_{add} = \frac{C}{n^3} \cdot [k_3 \pm k_4 \mp k_5 \cdot Z] \; , \qquad (19)$$

where $k_3$, $k_4$ and $k_5$ - integer factors, which values are given in Table V, and $C$ - the same constant, as in (13)-(16). The upper arithmetical sign in (19) corresponds to singlet states, and the lower sign – to triplet states.

**TABLE V. Factors $k_3$, $k_4$ and $k_5$.**

| State | $k_3$ | $k_4$ | $k_5$ |
|---|---|---|---|
| *1sns* | 2 | 4 | 3 |
| *1snp* | 2 | 3 | 2 |
| *1snd* | 0 | 0 | 0 |
| *1snf* | 0 | 0 | 0 |

It is easy to see, that the the VM$_1$ method gives so precise convergence with experiment for *1snd* and *1snf* states, that $E_{add}$ is equal to zero in this case. The calculations show that it is correct generally for all *1snl* states, at $l > 1$. It is one more proof that VM$_1$ is not one from many methods, but it is unique.

The general formula of total energy (11) for *1snl* states finally looks like

$$E_{He} = \varepsilon_1 \cdot Z_{e1}^2 + \varepsilon_2 \cdot \frac{Z_{e2}^2}{n^2} + \frac{C}{n^3} \cdot [k_3 \pm k_4 \mp k_5 \cdot Z] \qquad (20)$$

The data of calculation of energy $E_{He}$ under formula (14) under the supposition that $C = 1/9$ are given in Table IV.

## C. RESULTS AND COMMENTS

### 1. Physics

From a point of view of physics the formulas (13)-(16) and (19)-(20) lead to an unusual, i.e. to non-classical relation of energy of interaction of charged particles to the distance between them.

The interaction of some configurations of electrostatic charges in classical case leads to dependencies like

$$E \sim 1/r^t \; ,$$

where $E$ - energy of interaction, $r$ - distance between interacting configurations of electrostatic charges, $t = 1, 2, 3$ etc. In case $t = 1$ we have Coulomb's law, i.e. interaction of charges, and in case $t > 1$ we take into account dipolar, quadrupolar and other multypolar terms.

If we present principal quantum number $n$ as a radius of atom $r$ (remembering, that in hydrogen-like atoms $r \sim n^2$), it is easy to obtain for different parts of total energy of two-electron atom of dependency

$$E \sim 1/r^p \; ,$$

where $p = 1$ for terms $E_{He1}$, calculated by the VM$_1$ method that completely corresponds to the classical law of the Coulomb, and $p = 1/2$ and $3/2$ for additional energy $E_{add}$ respectively for intrashell and intershell states. Thus here, in difference from the classical case, firstly there are half-integer degrees of $r$, and secondly there is an additional interaction decreasing on a distance slower than the Coulomb force. This fact is the most difficult to explain and it attracts the most interest.

If we shall take into account the exchange degeneration in equation (5), i.e. if we shall take (4) instead of (3) for a common wave function of atom, it will cause to appearance of exchange integrals

$$\iint \psi_1(1)\psi_2(2)\frac{1}{r_{12}}\psi_2(1)\psi_1(2)dV_1dV_2 \qquad (21)$$

in addition to the coulomb integrals (7).

The appropriate exchange energy depends from $n$ as $E \sim 1/n^2$ for intrashell and $E \sim 1/n^3$ for intershell configurations. It can explain only part of numerical values of additional energy $E_{add}$ in both cases.

We can take into account also the "random" degeneration of the Coulomb field that results in the configuration interaction and in the mixing of states. However in this case the energy of electron-electron interaction also will be a sum of Coulomb and exchange integrals, and this sum also leads to dependences $E \sim 1/n^2$ for intrashell states. Thus consideration of various types of quantum degeneration does not give understanding of the origin of dependency $E_{add} \sim 1/n \sim 1/\sqrt{r}$ for intrashell states, though it allows to understand (not quantitatively, but at least qualitatively) the origin of dependency $E_{add} \sim 1/n^3 \sim 1/\sqrt{r^3}$ for intershell states.

Calculations have shown that the dependency $E_{add} \sim 1/n \sim 1/\sqrt{r}$ occurs only when both electrons have identical principal quantum numbers $n$. That fact can indicate resonant character of additional interaction. Moreover, it results in electron-electron attraction instead of an electrostatic repulsion and very strongly depends on the configuration of spin and orbital moment ( see Table II ) that makes it even less similar to electrostatic interaction, but similar to proton-proton interaction in the nucleus.

As one from versions, it is possible that the exotic relation $E \sim 1/n$ occurs because of the quantum effect – the wave-like nature of matter. In atom de-Broglie wavelength of electron $\lambda \sim r/n$ and we get for additional interaction $E \sim 1/\lambda \sim 1/n$ instead of the usual $E \sim 1/r \sim 1/n^2$ for electrostatic interaction.

Another possible explanation is that dependency $E \sim 1/n$ can be a consequence of the fact that the two-electron atom is a limit of a two-electron diatomic molecule at a zero nucleus-nucleus distance. It is known that the vibrational spectrum of a molecule is calculated like a spectrum of harmonic oscillator. I.e. actually, elastic forces are introduced at the calculation of a molecule in addition to Coulomb forces. It results in dependency $E \sim n$ for vibrational energy of a molecule where $n$ - a quantum number of harmonic oscillator. It is possible to assume that a similar additional elastic interaction is saved in the case of a two-electron atom. Then it would be possible to explain roughly the origin $E \sim 1/n$ by a certain superposition of dependencies $E \sim 1/n^2$ of hydrogen-like atom and $E \sim n$ of the harmonic oscillator. This superposition in case of coincidence of quantum numbers of atom and oscillator can result in $E \sim n \cdot \frac{1}{n^2} \sim \frac{1}{n}$.

## 2. Mathematics

From the mathematical point of view, the VM$_1$ method allows to divide the atom energy to Coulomb and non-Coulomb parts with very high accuracy. Thus, all Coulomb interaction is completely taken into account in calculation and non-Coulomb terms occur only in additional energy. It is correct for all types of states of helium-like atoms. Thus, if the non-classical interactions were absent in two-electron atom, the VM$_1$ method would give a precise analytical solution of a three-body problem in atomic physics. The simplicity of formulas (13)-(16) and (19) allows to hope that the analytical solution is possible also with the consideration of non-classical interactions. It would become possible after an evaluation of additional energy $E_{add}$ and constant $C$ from certain general principles. If it is possible, there will be no radical objections for obtaining the general analytical solution not only for two-electron atom (i.e. for a three-body problem), but also for atoms with any number of electrons (i.e. for N-body problem).

Let's add that if $C$ is the rational number and equals precisely 1/9, this is possible to present like $\frac{1}{9} = \frac{1}{3} \cdot \frac{1}{3}$, i.e. as result of interaction of certain additional charges of electrons. Similar fractional charges, multiple of 1/3, are encountered in microphysics in quarks and in macrophysics in a fractional quantum Hall effect.

It is possible also that $C$ actually is an irrational number. Let's consider, as an example, doubly excited state $nsns$(1S), described by formula (16). In this case, value $n$ at which the energy of non-Coulomb electron-electron attraction $E_{add}$ begins to exceed the energy of Coulomb repulsion $E_{rep}$, almost precisely equals value $n$, at which the radius of hydrogen-like atom begins to exceed the minimum wavelength of a spectrum of radiation of the given atom which corresponds to energy of ionization. If it is not a simple accidental coincidence, then it is possible to record instead of (16)

$$E_{add} = -E_{rep} \cdot n \cdot \sqrt{\alpha} \; , \qquad (22)$$

where $E_{rep}$ equals Coulomb integral (7), and $\alpha$ - fine-structure constant that is irrational and approximately equals $\frac{1}{137}$. It is obvious that $E_{add} > - E_{rep}$ at $n > \frac{1}{\sqrt{\alpha}}$, i.e. at $n > 11$ taking into account that $n$ is an integer. Thus $C$ can be determined like

$$C = -E_{rep} \cdot \frac{n^2}{Z} \cdot \sqrt{\alpha} \; . \qquad (23)$$

In case of a ground state 1s1s, $n = 1$, $E_{rep}$ = -1,25·$Z$ and $C$ approximately equals 0,107. If $n$ increases, the value $E_{rep} \cdot n^2$ slightly decreases and aims to limit $E_{rep} \cdot n^2 = -1,20 \cdot Z$ that corresponds to a limit of 0,100 for $C$. Obtained values for $C$ correspond to the data of Table I taking into account errors of measurements.

### III. INTER-ATOMIC INTERACTIONS

#### A. Molecules

In case of a molecule of hydrogen $H_2$ Coulomb integral (7), giving the correction to zero approximation, is noted in accordance with [16] like

$$\iint \psi_1^2 \cdot (\frac{1}{r_{ab}} + \frac{1}{r_{a2}} + \frac{1}{r_{b1}} + \frac{1}{r_{12}}) \cdot \psi_2^2 dV_1 dV_2 \qquad , \qquad (24)$$

where $r_{ab}$ - distance between nucleus $a$ and $b$, $r_{12}$ - distance between 1-st and 2-nd electrons which in zero approximation are located close to $a$ and $b$ respectively, $r_{a2}$ and $r_{b1}$ - distance between the nucleus and «foreign» electron, and $\psi_1$ and $\psi_2$ - hydrogen-like wave functions.

For a ground state of $H_2$ the calculation of the Coulomb integral (24) according to [16] looks (in a.e.) like

$$\frac{e^{-2\rho}}{\rho} \cdot (1 + \frac{5}{8}\rho - \frac{3}{4}\rho^2 - \frac{1}{6}\rho^3) \qquad , \qquad (25)$$

where $\rho = \frac{r_{ab}}{r}$ , where $r$ - radius of atom of hydrogen.

If we substitute to this formula a value known from the experiment, $\rho$ = 1,4006 for ground state $H_2$, the numerical value (25) will be equal 0,00233 a.e. or 0,0047 Ry. If we compare it to the known value of a dissociation energy of $H_2$ 0,3323 Ry, we come to a conclusion that the classical Coulomb interaction results almost in zero energy of chemical bond. I.e. almost all energy of bond $H_2$ has a non-classical origin. This outcome was expected since the homopolar chemical bond was explained only in quantum mechanics. Even heteropolar ionic bonds eventually are non-classical because negative ions included in their structure cannot be described using only classical interactions.

Thus we can assume with a sufficient degree of accuracy that all energy of the chemical bond is non-classical, i.e. additional to the Coulomb interaction. It gives us a simple empirical way of determination of non-classical energy of interatomic interaction by experimental energies of interatomic bonds. It is different from the case of intraatomic interaction when the calculations were necessary for the separation of classical energy from non-classical.

Let's assume that all influence of internal electrons of atoms on the chemical bond is reduced only to shielding a charge of the nucleus for external valence electrons. Then the ground states of molecules $Li_2$, $Na_2$, $K_2$, $Rb_2$ can be presented as doubly excited states of the molecule $H_2$.

Values of energy of the chemical bond $E_b$ of molecules of elements of the first group of the periodic table, general quantum numbers $n$ of valence electrons and product $E_b \cdot n$ are given in table VI. Experimental values of energy of bonds are taken from [17] and transformed in Ry.

**TABLE VI. The data of energy of bonds diatomic molecules of the elements of the 1-st group.**

| Molecule | $n$ | $E_b$ (kJ/mole) | $E_b$ (Ry) | $E_b \cdot n$ (Ry) |
|:---:|:---:|:---:|:---:|:---:|
| $H_2$ | 1 | 436 | 0,3323 | 0,3323 |
| $Li_2$ | 2 | 101,7 | 0,0775 | 0,1550 |
| $Na_2$ | 3 | 73 | 0,0556 | 0,1669 |
| $K_2$ | 4 | 54 | 0,0412 | 0,1646 |
| $Rb_2$ | 5 | 49 | 0,0373 | 0,1867 |

It is easy to see that for $Li_2$, $Na_2$ and $K_2$ the dependency $E_b \sim 1/n$ is explicitly observed. It is possible to explain the deviation in $Rb_2$ by increasing the number of internal electrons and by a very large nucleus charge that increases the role of relativistic effects. It is possible to explain the deviation in case $H_2$, as well as in the case of a ground state of atom He (see item II B) that the 1s1s is the lowest possible state of the configuration $nsns$.

It is possible to record the general formula for energy of bonds $Li_2$, $Na_2$ and $K_2$ approximately like

$$E_b = \frac{3}{2} \cdot \frac{C}{n} , \qquad (26)$$

where $C$ - same constant as in item II.

For $H_2$ the formula for energy of the bond can be written like

$$E_b = 3 \cdot C . \qquad (27)$$

Let's note that the energy of bond $H_2$ with a large degree of accuracy equals 1/3 (in Ry) if $C$ equals exact 1/9.

By analogy to Table VI, we shall make Table V for diatomic molecules of the elements of the 5,6 and 7 groups of the periodic table. Experimental values of energy of bonds are taken from [17].

**TABLE VII. The data of energy of bonds diatomic molecules of the elements of the 5,6 and 7-th groups**

| Molecule | $n$ | $E_b$ (kJ/mole) | $E_b$ (Ry) | $E_b \cdot n$ (Ry) |
|:---:|:---:|:---:|:---:|:---:|
| **5 group** | | | | |
| $N_2$ | 2 | 945,3 | 0,7205 | 1,4410 |
| $P_2$ | 3 | 489,1 | 0,3728 | 1,1184 |
| **As₂** | 4 | 385 | 0,2935 | 1,1738 |
| **Sb₂** | 5 | 323 | 0,2462 | 1,2310 |
| **6 group** | | | | |
| **O₂** | 2 | 498,4 | 0,3799 | 0,7598 |
| **S₂** | 3 | 425,5 | 0,3243 | 0,9729 |
| **Se₂** | 4 | 309 | 0,2355 | 0,9421 |
| **Te₂** | 5 | 259 | 0,1974 | 0,9870 |
| **7 group** | | | | |
| **F₂** | 2 | 159 | 0,1212 | 0,2424 |
| **Cl₂** | 3 | 239,2 | 0,1823 | 0,5470 |
| **Br₂** | 4 | 201 | 0,1532 | 0,6128 |
| **I₂** | 5 | 151,1 | 0,1152 | 0,5758 |

It's obvious that one can observe a dependency $E_b \sim 1/n$, and it is possible to explain deviations (as in the case of the first group) on the one hand by a special rule for the lowest possible states and with another - by increasing the influence of internal electrons and relativistic effects.

## B. Crystals

In Table VIII by analogy with Tables VI and VII the data for energy of bonds (i.e. energy of sublimation of lattice) $E_b$ are given for metal crystals Li, Na, K, Rb that are elementary cases of crystals in general and metals in particular. Experimental values are taken from [18].

**TABLE VIII. Energy of bonds lattices of elements of the 1-st group**

| | $n$ | $E_b$ (kJ/mole) | $E_b$ (Ry) | $E_b \cdot n$ (Ry) |
|---|---|---|---|---|
| **Li** | 2 | 163 | 0,1242 | 0,2485 |
| **Na** | 3 | 109 | 0,0831 | 0,2492 |
| **K** | 4 | 84 | 0,0640 | 0,2561 |
| **Rb** | 5 | 79 | 0,0602 | 0,3011 |

The general formula of energy of bonds of lattice for these elements can be written down approximately like

$$E_b = \frac{9}{4} \cdot \frac{C}{n},$$ (28)

where $C$ – same constant as in item II.

Thus on all levels of a chemical (not nuclear) matter - from two-electron atoms up to molecules and crystals - we find proofs that there is a non-classical attraction between electrons for which the dependency $E \sim 1/n$ is observed when the interacting electrons have identical principal quantum numbers $n$. Moreover, numerical values of additional non-classical energy express by simple rational numbers through a constant $C$ that also occurs on all levels of the structure of chemical matter and probably has a fundamental character.

## IV. PRACTICAL CONSEQUENCES

From a practical point of view it is interesting that in the case of $nl_1nl_2$ states, since some value $n$, the usual Coulomb electron-electron repulsion (decreasing as $1/n^2$) will become less than additional non-Coulomb attraction (decreasing as $1/n$). It can result in the macroscopic case in the joining of electrons in certain stable or metastable structures just like protons are integrated in the nucleus. Similar processes could happen spontaneously in nature. It is possible that similar effects could explain at least some of the anomalous plasma-like effects observed sometimes in the atmosphere and in the ionosphere, such as ball lightning etc.

To receive similar new states of matter in the experiment, it is necessary that electrons of substance are excited synchronously, i.e. have identical energy and identical values $n$ in each instant. To the present moment not much of similar (doubly exited) states has been obtained even for two-electron atoms. They're even less known for molecules. Moreover both in the case of atoms, and in the case of molecules the values $n$ don't yet reach the value at which the electron-electron attraction exceeds repulsion. For example, for $nsns$ states of two-electron atoms only $n = 3$ is reached in the experiment, but $n > 11$ is required (see item II C).

In the case of the macroscopic quantity of matter the problem of synchronous excitation of electrons up to the maximum large $n$ was not even posed yet, though technically it is possible.

Let us also remind that the explanation of superconductivity involves the appearance of electron-electron attraction (its origin and physical nature doesn't matter), which exceeds the Coulomb repulsion under certain conditions. It makes probable bose-einstein condensation of synchronously excited electrons both in the case of separate atoms, and in the case of macroscopic quantity of matter from that moment when additional non-Coulomb electron-electron attraction will begin to exceed Coulomb repulsion. The transition in a superconducting state is possible for fixing even in the case of separate atoms. If our suppositions are correct, the atoms can be transformed into an ideal diamagnetic at some critical value $n$ of synchronously excited electrons. Similar superconductivity already could be named super-high-temperature.

## V. CONCLUSION

The approach based on the separation of total energy of multyelectron systems into classical Coulomb and non-classical additional parts, allows on the one hand to simplify calculations and on the other to see very simple and very interesting relations that were not visible at the use of more complex methods.

The most interesting aspect is that on all levels of chemical (not nuclear) matter - from two-electron atoms up to molecules and crystals - we find non-classical electron-electron attraction and the dependency $E \sim 1/n$ for energy of this attraction when the interacting electrons have identical principal quantum numbers $n$. Moreover, numerical values of additional non-classical energy expressed by simple rational numbers through a constant $C$ that also occurs on all levels of chemical matter and probably has a fundamental character.

The most important practical consequence is the discovery that under certain conditions the electron-electron attraction can exceed the electron-electron repulsion. It can result in the existence of ordered structures of a new type in a special-way exited substance.

The author of the present work hopes that the relations obtained by him, as well as practical conclusions and the predictions of new physical effects can interest theorists and experimenters working in the field of physics of atom and molecules, physics of condensed matter and physics of plasma.

## REFERENCE

[1]   G. Tanner, K. Richter and J.M. Rost,  Rev.Mod.Phys. **72**, 497 (2000)
[2]   H.A. Bethe and E.E. Salpeter,  *Quantum Mechanics of One- and Two- Electron Atoms* (New York, 1957)
[3]   J. Emsley  *The Elements* (Clarendon press, Oxford 1991)
[4]   NIST Standard Reference Database
      http://www.physics.nist.gov/PhysRefData/ASD/index.html
[5]   S. Buckman and C. Clark, Rev.Mod.Phys. **66**, 539 (1994)
[6]   P.J. Hicks and J. Comer,  J. Phys. B: At. Mol. Phys. **8**, 1866 (1975)
[7]   M. Rodbro, R. Bruch and P. Bisgaard,  J. Phys. B: At. Mol. Phys. **12**, 2413 (1979)
[8]   H.A. Sakaue et al.,  J. Phys. B: At. Mol. Opt. Phys. **24**,3787 (1991)
[9]   K. Iemura et al., Phys.Rev. A **64**, 062709 (2001)

[10]  M. Mack, J. Nijland , P. Straten, A. Niehaus and R. Morgenstern, Phys. Rev. A **39**, 3846 (1989)

[11]  D.H. Oza, P. Benoit-Cattin, A. Bordenave-Montesquieu, M. Boudjema and A. Gleizes,  J. Phys. B: At. Mol. Opt. Phys. **21**, L131 (1988)

[12]  M. Domke, K. Schulz, G. Remmers, G. Kaindl and D. Wintgen, Phys. Rev. A **53**, 1424 (1996)

[13]  H.G. Berry, J. Desesquelles and M. Dufay, Phys. Rev. A **6**, 600 (1972)

[14]  S.J. Brotton, S. Cvejanovic, F. Currell, N. Bowring and F.H. Read, Phys. Rev. A **55**, 318 (1997)

[15]  S.W. J Scully et al., J. Phys. B: At. Mol. Opt. Phys. **39**, 3957 (2006)

[16]  A. Sommerfeld, *Atombau und Spectrallinen* (1939, in german)

[17]  V.A. Rabinovich, Z.Y. Havin, *Short chemical handbook* (Moscow, 1978, in russian)

[18]  G.I. Novikov, *Foundations of general chemistry* (Moscow, 1988, in russian)